\documentclass[prb,twocolumn,floatfix]{revtex4}
\usepackage{amsfonts}
\usepackage{amssymb}
\usepackage{amsmath}
\usepackage[dvips]{graphicx}

\begin{document}

\title{Contact resistance and shot noise in graphene transistors}
\author{J. Cayssol}
\affiliation{CPMOH, UMR 5798, Universit\'{e} de Bordeaux,\\
33405 Talence, France}
\author{B. Huard}
\thanks{Present address: Laboratoire Pierre Aigrain, D\'epartement de
Physique de l'Ecole Normale Sup\'erieure\\
24 rue Lhomond, 75231 Paris Cedex 05, France}
\author{D. Goldhaber-Gordon}
\affiliation{Stanford University, Department of Physics, Stanford, California, USA }

\begin{abstract}
Potential steps naturally develop in graphene near metallic contacts. We
investigate the influence of these steps on the transport in graphene Field
Effect Transistors. We give simple expressions to estimate the
voltage-dependent contribution of the contacts to the total resistance and
noise in the diffusive and ballistic regimes.
\end{abstract}

\maketitle

Graphene's distinctive band structure gives rise to exciting new transport
properties and promising applications for carbon-based electronics \cite%
{avouris07,castroneto07,beenakker08}. When measuring the conductance or
current noise in a nanotube or a sheet of graphene, the properties of the
contacts can matter as much as the electronic structure of the nanotube or
graphene itself. In semiconducting nanotubes or graphene nanoribbons, it is
known that Schottky barriers develop at the metallic contacts \cite%
{Heinze02,Javey05}. Charge transfer between a metal and a wide graphene
strip induces potential steps whose shape may differ strongly from usual
Schottky barriers due to the semimetallic and two-dimensional nature of
graphene. The existence of such metal-induced potential steps was inferred
experimentally from the transport properties of a graphene strip with
various contact geometries \cite{huard08}. More direct evidence for these
steps comes from optical mapping of the potential landscape across a
graphene device \cite{EduardoLee08}. Recent theoretical work on
graphene-metal interfaces has been performed within the atomistic
tight-binding theory \cite{schomerus07,blanter07}.

In this paper, the conductance and the shot noise of graphene Field Effect
Transistors (gFETs) with extended contacts are derived using the Dirac
Hamiltonian for graphene. Near a single contact, we assume that the Fermi
energy in graphene varies monotonically over a characteristic length $d$,
and we solve the corresponding scattering problem exactly. If the transport
is ballistic between both contacts, we predict oscillations of the noise and
conductance as the charge density is increased in the sheet. When the
density exceeds the one under the contact, the noise minima might be zero
and correspond to perfect transmission between the contacts. Such
realizations of a noiseless gFET are caused by Fabry-P\'{e}rot resonances
and require low doping by the contacts. In the diffusive regime, we show how
the total resistance and Fano Factor of the whole gFET depend upon the
contact resistance and Fano factor of each contact, which is relevant in
interpreting recent experiments on shot noise in non-suspended graphene \cite%
{dicarlo08,danneau08}. 
\begin{figure}[tbph]
\begin{center}
\includegraphics[width=8.5cm]{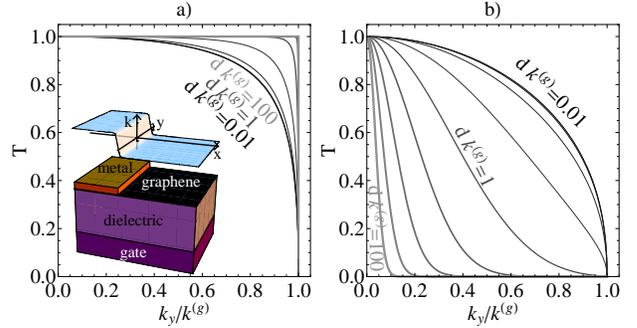}
\end{center}
\par
\vspace*{-0.5cm}
\caption{a) Transmission probability $\mathcal{T}_{\mathrm{step}}$ across
the metal/graphene interface as a function of transverse momentum $k_{y}$
for several lengths $d$ of the potential rise. The ratio between Fermi
wavevectors in bulk graphene and below the metal is set to $%
k_{F}^{(m)}/k_{F}^{(g)}=3$ (unipolar contact). A schematic of the device and
corresponding Fermi wavevector profile $k_{F}(x)$ are shown in the inset. b)
same curves for a bipolar contact with $k_{F}^{(m)}/k_{F}^{(g)}=-3$.}
\label{figureT}
\end{figure}

Before analyzing the gFETs' properties, it is useful to investigate
transport across a single graphene-metal contact. We thus consider that a
metal electrode covers the left half-plane ($x<0$) of an infinite graphene
layer. We assume that the metal coating simply shifts the Fermi level of the
graphene underneath while preserving its pristine Dirac cones \cite%
{giovannetti08}. Far from the contact, the type ($n$ or $p$) and density of
charge carriers in the right half-plane ($x>0$) are tuned by a distant
metallic gate. A continuous Fermi wave vector profile $k_{F}(x)$ must
therefore develop near the contact edge to match the asymptotic values $%
k_{F}(-\infty )=k_{F}^{(m)}$ under the metal, and $k_{F}(+\infty
)=k_{F}^{(g)}$ in bare graphene. The dynamics of the massless fermions can
be safely described by the single-valley two-dimensional Dirac Hamiltonian, 
\begin{equation}
H=-i\hbar v_{F}\left[ \sigma _{x}\partial _{x}+\sigma _{y}\partial
_{y}-ik_{F}(x)\mathbf{1}\right] ,
\end{equation}%
which is valid if the potential step is smooth on the scale of the lattice
constant. Here $v_{F}$ is the Fermi velocity and 
\begin{equation}
k_{F}(x)=k_{F}^{(m)}+\frac{k_{F}^{(g)}-k_{F}^{(m)}}{e^{-x/d}+1}
\label{potstep}
\end{equation}%
our simple model for the space-dependent Fermi wave vector (Fig.~\ref%
{figureT} inset). The Pauli matrices ($\sigma _{x},\sigma _{y}$,$\sigma _{z}$%
) operate on spinors $\Psi (x,y)$ whose components are the electron
amplitudes associated with each sublattice of the honeycomb carbon crystal.
It is worthy to note that a complete treatment of nonlinear screening and
disorder near the contact should lead to a more complicated profile in the
bipolar $n$-$p$ case \cite{fogler08,stander08}. Nevertheless this can be
accounted for by noting that our phenomenological $d$ will in fact depend on 
$k_{F}^{(m)}$ and $k_{F}^{(g)}$ in a way that may ultimately be calculated.

We now proceed to the derivation of simple formulas relating the transport
properties (as functions of $k_{F}^{(g)}$) to the parameters $d$ and $%
k_{F}^{(m)}$ characterizing the contacts. Here we consider a wide enough
graphene strip to neglect edge effects \cite{footnote}. Then translational
invariance parallel to the junction (along the $y$ axis) implies that the
transverse wavevector $k_{y}$ is conserved, and all spinor wavefunctions
take the form $\Psi (x,y)=\Phi (x)e^{ik_{y}y}$ where $\Phi (x)=~^{T}(\Phi
_{1}(x),\Phi _{2}(x))$. Aiming to determine the low temperature transport,
we solve the Dirac equation at the Fermi level ($H\Psi (x,y)=0$) which
reduces to the one-dimensional equation%
\begin{equation}
\partial _{x}\Phi (x)=(k_{y}\sigma _{z}+ik_{F}(x)\sigma _{x})\Phi (x).
\label{dirac}
\end{equation}%
This equation can be decoupled using the symmetric/antisymmetric
combinations of components $f_{\pm }(x)=\Phi _{1}(x)\pm \Phi _{2}(x)$ which
obey the following scalar differential equations: 
\begin{equation}
f_{\alpha }^{\prime \prime }+(k_{F}^{2}(x)-k_{y}^{2}-i\alpha k_{F}^{^{\prime
}}(x))f_{\alpha }=0,  \label{f1}
\end{equation}%
where $\alpha =\pm $. In the asymptotic regions $\left\vert x\right\vert \gg
d$, the solutions are plane waves $f_{\alpha }(x)=e^{\pm ik_{x}x}$ which can
be either exponentially damped or oscillatory depending on the sign of $%
k_{x}^{2}=k_{F}^{2}-k_{y}^{2}$. In order to find the transmission across the
potential step Eq.~(\ref{potstep}), we now construct a scattering state
containing a single oscillatory outgoing charge state (with $%
|k_{y}|<|k_{F}^{(g)}|$) in the region $x\rightarrow +\infty $, namely 
\begin{equation}
f_{\alpha }(x)\sim e^{ik_{x}^{(g)}x}\mbox{ at }x\rightarrow +\infty .
\label{fplusinf}
\end{equation}%
Here $k_{x}^{(g)}=s_{g}\sqrt{(k_{F}^{(g)})^{2}-k_{y}^{2}}$ is the
longitudinal momentum whose direction depends on the band index $s_{g}=%
\mathrm{sign}(k_{F}^{(g)})$ far on the right side. In Appendix A, we show
that this asymptotic condition completely determines the solution of the
Dirac equation $H\Psi (x,y)=0$ on the whole $x$-axis. In particular, on the
left side the wave consists of a superposition%
\begin{equation}
f_{\alpha }(x)\sim f_{\alpha }^{(inc)}(x)+f_{\alpha }^{(ref)}(x)\mbox{ at }%
x\rightarrow -\infty ,  \label{fmoinsinf}
\end{equation}%
of an incoming 
\begin{equation}
f_{\alpha }^{(inc)}(x)=\frac{\Gamma (1-2ik_{x}^{(g)}d)\Gamma
(-2ik_{x}^{(m)}d)}{\Gamma (i\kappa ^{--}d)\Gamma (1-i\kappa ^{++}d)}%
e^{ik_{x}^{(m)}x},  \label{fmoinsinf1}
\end{equation}%
and a reflected charge carrier 
\begin{equation}
f_{\alpha }^{(ref)}=\frac{\Gamma (1-2ik_{x}^{(g)}d)\Gamma (2ik_{x}^{(m)}d)}{%
\Gamma (i\kappa ^{-+}d)\Gamma (1-i\kappa ^{+-}d)}e^{-ik_{x}^{(m)}x}.
\label{fmoinsinf2}
\end{equation}%
Here $k_{x}^{(m)}=s_{m}\sqrt{(k_{F}^{(m)})^{2}-k_{y}^{2}}$ and $s_{m}=%
\mathrm{sign}(k_{F}^{(m)})$ indicates whether graphene is $n$- or $p$-doped
underneath the metal. The Euler Gamma function is denoted $\Gamma (z)$ and
we have introduced the momenta $\kappa ^{\rho \sigma
}=k_{F}^{(g)}-k_{F}^{(m)}+\rho k_{x}^{(g)}+\sigma k_{x}^{(m)}$, with $\rho
,\sigma =\pm 1$. The corresponding reflection probability is simply given by%
\begin{equation}
\mathfrak{R}_{\mathrm{step}}=\left\vert \frac{\Gamma (1-i\kappa
^{++}d)\Gamma (i\kappa ^{--}d)}{\Gamma (1-i\kappa ^{+-}d)\Gamma (i\kappa
^{-+}d)}\right\vert ^{2}  \label{RGamma}
\end{equation}%
when all the waves are propagating, namely for $\left\vert k_{y}\right\vert <%
\mathrm{min}(|k_{F}^{(g)}|,|k_{F}^{(m)}|)$. Finally, a remarkably simple
formula is obtained for the reflection coefficient of Dirac fermions across
the potential step Eq.~(\ref{potstep}): 
\begin{equation}
\mathfrak{R}_{\mathrm{step}}=\frac{\sinh (\pi d\kappa ^{+-})}{\sinh (\pi
d\kappa ^{++})}\frac{\sinh (\pi d\kappa ^{-+})}{\sinh (\pi d\kappa ^{--})}.
\label{R1}
\end{equation}%
This expression is valid for any $d$ and step polarity, matching the known
limits for transport across smooth \cite{cheianov06} and abrupt ($%
d\rightarrow 0$) steps \cite{katsnelson06}, and interpolating between those
limits. The expression~(\ref{R1}) is reminiscent of the reflection
coefficient of a non-relativistic massive scalar particle, $\mathfrak{R}%
=\sinh \left( \pi d\left\vert k_{x}^{(g)}-k_{x}^{(m)}\right\vert \right)
/\sinh \left( \pi d\left\vert k_{x}^{(g)}+k_{x}^{(m)}\right\vert \right) $,
obtained by solving the Schr\"{o}dinger equation in a similar potential
landscape \cite{landauT3}. The richer structure of Eq.~(\ref{R1}) is
associated with the Dirac nature of carriers in graphene. In particular, it
indicates the absence of backscattering at normal incidence ($%
k_{x}^{(m,g)}=k_{F}^{(m,g)}$) for any height and width of the potential
step, which is related to the orthogonality of incoming and reflected spinor
states \cite{ando98}. 
\begin{figure}[tbph]
\begin{center}
\includegraphics[width=9.cm]{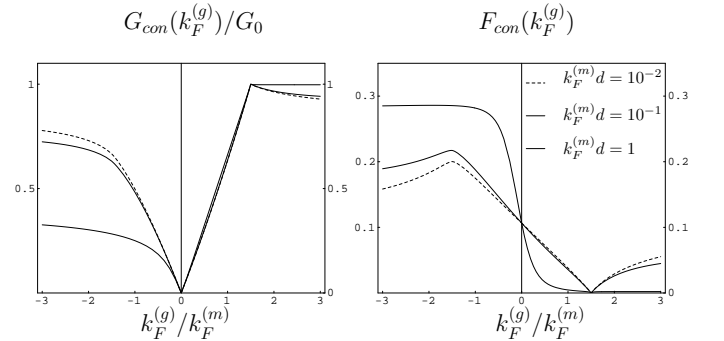}
\end{center}
\par
\vspace*{-0.7cm}
\caption{Contact conductance (left) and Fano factor (right) associated with
a single potential step as functions of the Fermi wavevector in bulk
graphene $k_{F}^{(g)}$ for several values of the dimensionless parameter $%
k_{F}^{(m)}d$. The maximal value of the conductance is given by $%
G_{0}=4e^{2}N/h$ where $N=k_{F}^{(m)}W/\protect\pi $. }
\label{figure2a}
\end{figure}
As with standard impedance matching, the transmission $\mathcal{T}_{\mathrm{%
step}}\equiv 1-\mathfrak{R}_{\mathrm{step}}$ of the unipolar contact tends
toward unity when the distance $d$ is increased (Fig.~\ref{figureT}). In
contrast, for bipolar steps, the $k_{y}$-dependent transmission $\mathcal{T}%
_{\mathrm{step}}$ goes from a broad curve at small $d$ to a sharp peak
around $k_{y}=0$ at large $d$ (Fig.~\ref{figureT}). Besides, the bipolar
transmission counter-intuitively increases when the potential barrier height
is increased. Although the potential step originates from charge transfer
between a metal and graphene, these scattering properties are similar to the
relativistic Klein tunneling \cite{klein29} for which evidence is mounting
in the context of transport through potential barriers created by local
gates \cite%
{huard07,williams07,ozyilmaz07,oostinga07,gorbachev08,liu08,stander08}.

From the transmission probability $\mathcal{T}_{\mathrm{step}}$, the
conductance and the Fano factor of a single contact are given respectively
by 
\begin{equation}
G_{\mathrm{con}}(k_{F}^{(g)})=\frac{4e^{2}}{h}\frac{W}{2\pi }%
\int_{-k_{F}^{(m)}}^{k_{F}^{(m)}}\mathcal{T}_{\mathrm{step}}\mathrm{d}k_{y},
\label{Rconexpression}
\end{equation}%
and 
\begin{equation*}
F_{\mathrm{con}}(k_{F}^{(g)})=\int_{-k_{F}^{(m)}}^{k_{F}^{(m)}}\mathrm{d}%
k_{y}\mathcal{T}_{\mathrm{step}}(1-\mathcal{T}_{\mathrm{step}%
})/\int_{-k_{F}^{(m)}}^{k_{F}^{(m)}}\mathrm{d}k_{y}\mathcal{T}_{\mathrm{step}%
},
\end{equation*}%
where $W$ is the width of the graphene strip along $y$ and $k_{F}^{(g)}$ is
related to the asymptotic density $n^{(g)}$ at $x\gg d$ by the relation $%
k_{F}^{(g)}=\sqrt{\pi n^{(g)}}$. These quantities are strongly sensitive to
the nature and density of charge carriers (Fig.~\ref{figure2a}). The bipolar
contact is clearly more resistive and noisier than the unipolar one, and
this unipolar/bipolar asymmetry becomes more pronounced for smoother
contacts (Fig.~\ref{figure2a}). In the limit of very smooth potential steps $%
(k_{F}^{(m)}d\rightarrow \infty )$, the Fano factor vanishes in the unipolar
case whereas it saturates to a finite value $F_{\mathrm{con}%
}(k_{F}^{(g)}<0)=1-2^{-1/2}$ in the bipolar case (in agreement with \cite%
{cheianov06}). As expected a single contact becomes noiseless when the
potential step vanishes ($k_{F}^{(g)}=k_{F}^{(m)}$). 
\begin{figure}[tbph]
\begin{center}
\includegraphics[width=7.5cm]{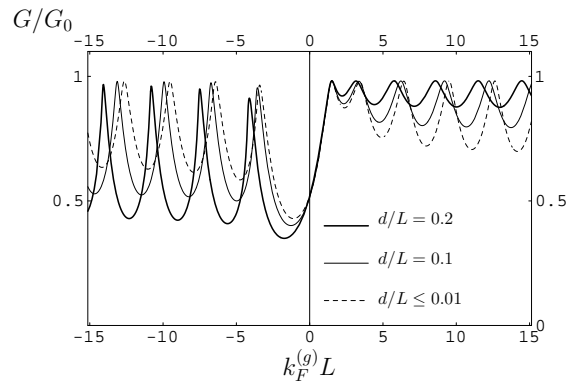}
\end{center}
\par
\vspace*{-0.5cm}
\caption{Conductance $G$ of a ballistic Fabry-P\'{e}rot cavity as a function
of the dimensionless parameter $k_{F}^{(g)}L$. The maximal conductance ($%
G_{0}=4e^{2}N/h$) is controlled by the number of available propagating modes
in the leads $N=k_{F}^{(m)}W/\protect\pi $. Here $k_{F}^{(m)}L=1.5$, which
means $L=27$~nm for a density $n^{(m)}=10^{11}~\mathrm{cm}^{-2}$, and the
aspect ratio $W/L$ is large (in practice $W/L\gtrsim 4$ is sufficient 
\protect\cite{footnote}). For definiteness, we have assumed that the
graphene is $n$-doped underneath the metal ($k_{F}^{(m)}>0$). In this case,
the minimal conductivity is reached at a nonzero negative density in the
central channel. For other metals that produce $p$-doped graphene after
charge transfert, one should obtain the symmetrical curves (with respect to
vertical axis $k_{F}^{(g)}L=0$) of the ones presented here, and the shift of
the minimal conductivity would occur at positive density. }
\label{figureGbal}
\end{figure}
In the sequel of the paper, we use the result Eq.~(\ref{R1}) to investigate
the effect of the contact potential steps \cite{huard08,EduardoLee08} on the
conductance and noise properties of gFETs.

Recently, suspended gFETs have been achieved, resulting in increased
mobility of the two-dimensional electron gas \cite%
{bolotin08,andrei08,young08}. In such devices electronic motion might become
ballistic between source and drain. The whole structure can be described by
two symmetric steps similar to Eq.~(\ref{potstep}) and separated by a
distance $L$, the so-called Wood-Saxon potential Eq.~(\ref{wood-saxon}). The
single channel transmission exhibits Fabry-P\'{e}rot-like resonances as the
gate voltage varies. We calculate the transmission probability using a
mapping between the problem of massless Dirac fermions in graphene and the
one of massive Dirac fermions in the one-dimensional Wood-Saxon potential 
\cite{kennedy02}. The contrast of the whole interference pattern is
controlled by the reflection probability $\mathfrak{R}_{\mathrm{step}%
}(k_{y}) $ which depends strongly on the distance $d$, as discussed above,
see Eq.~(\ref{R1}).

In the large gate voltage regime, $\left\vert k_{F}^{(g)}\right\vert
>\left\vert k_{F}^{(m)}\right\vert $, the wavefunctions are oscillatory for
all transverse channels, and the corresponding transmission across the whole
device can be approximated by 
\begin{equation}
\mathcal{T}_{\mathrm{ball}}(k_{y})\approx \left( 1+\frac{4\mathfrak{R}_{%
\mathrm{step}}(k_{y})}{(1-\mathfrak{R}_{\mathrm{step}}(k_{y}))^{2}}\sin
^{2}(k_{x}^{(g)}L)\right) ^{-1}  \label{T1}
\end{equation}%
when neglecting the corrections of order $d$ to the effective size of the
cavity. The shape of the cavity enters this formula through the reflection
probability $\mathfrak{R}_{\mathrm{step}}(k_{y})$\textbf{. } The actual
position of the conductance peaks (Fig.~\ref{figureGbal}) is modified
because the effective width of the cavity changes with $d$. In this regime,
the conductance of realistic cavities is globally smaller (Fig.~\ref%
{figureGbal}) than the prediction of the square well-model ($d=0$) with
infinite doping below the electrodes ($k_{F}^{(m)}=\infty $) \cite%
{tworzydlo06}. Indeed the maximal conductance ($4e^{2}N/h$) is controlled by
the number of available propagating modes in the leads $N=k_{F}^{(m)}W/\pi $
as soon as the doping of the two-dimensional electron gas in the middle part
of the gFET exceeds that underneath the source and drain contacts. At large
absolute gate voltages ($k_{F}^{(g)}<-k_{F}^{(m)})$, the bipolar conductance
exhibits strong oscillations whose contrast increases as the steps become
smoother (Fig.~\ref{figureGbal}). In contrast at $k_{F}^{(g)}>k_{F}^{(m)}$,
the unipolar cavity becomes fully transparent in the large $d$ limit wherein
the oscillations are lost. Similar fringes have already been observed
recently in short Fabry-P\'{e}rot devices created by local gating \cite%
{young08}. We suggest making contacts with a metal which dopes the graphene
very lightly in order to observe the Fabry-P\'{e}rot interferences with the
highest resolution in bipolar cavities.

We now discuss the more usual low gate voltage regime $\left\vert
k_{F}^{(g)}\right\vert <\left\vert k_{F}^{(m)}\right\vert $, wherein the
conductance is determined by both evanescent and propagating modes. Besides
being nonuniversal, the minimal conductance is reached at nonzero carrier
density $n_{\text{min}}^{(g)}\neq 0$. Up to now, the minimum conductance was
predicted to occur at a non-zero gate voltage (due to charge impurities) but
at zero average density\cite{chen08}. Here, we predict that the presence of
metallic contacts can shift this minimum to non-zero charge density, even
without impurities. 
\begin{figure}[tbph]
\begin{center}
\includegraphics[width=7.5cm]{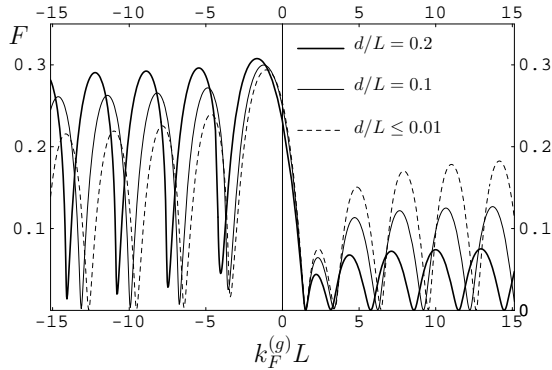}
\end{center}
\par
\vspace*{-0.7cm}
\caption{Fano factor for a ballistic graphene Fabry-P\'{e}rot cavity as a
function of the dimensionless parameter $k_{F}^{(g)}L$ for various values of 
$d/L$. Same parameters as in Fig. 3, in particular $k_{F}^{(m)}L=1.5$ while
the aspect ratio $W/L$ is large \protect\cite{footnote}, namely $W/L\gtrsim
4 $. }
\label{figureFbal}
\end{figure}
Besides, the Fabry-P\'{e}rot interferences should also induce signatures on
the noise of clean gFETs. Here we demonstrate that the Fano factor of a
ballistic gFET, 
\begin{equation}
F_{\mathrm{ball}}(k_{F}^{(g)})=\frac{\int \mathrm{d}k_{y}\mathcal{T}_{%
\mathrm{ball}}(1-\mathcal{T}_{\mathrm{ball}})}{\int \mathrm{d}k_{y}\mathcal{T%
}_{\mathrm{ball}}},
\end{equation}%
should exhibit spectacular oscillations as a function of the dimensionless
parameter $k_{F}^{(g)}L$ (Fig.~\ref{figureFbal}) for cavities operated at
gate voltages yielding $\left\vert k_{F}^{(g)}\right\vert >k_{F}^{(m)}$. In
particular for $k_{F}^{(g)}>k_{F}^{(m)}$, the Fano factor shows successive
nodes at the gate voltages corresponding to peaks of optimal conductance
(Fig.~\ref{figureGbal}). On the bipolar side $k_{F}^{(g)}<-k_{F}^{(m)}$, the
nodes are replaced by local minima of $F_{\mathrm{ball}}$ corresponding to
local maxima of the conductance. These oscillations have not been
investigated either experimentally or theoretically. Early theoretical works
assumed infinite doping by the contacts \cite{tworzydlo06}.

We now consider the situation, relevant to non-suspended gFETs, where the
carrier motion is diffusive between the source and the drain. Assuming
phase-incoherent transport, the total resistance of the gFET is simply given
by $R_{\mathrm{dif}}=R_{\mathrm{sheet}}(k_{F}^{(g)})+2R_{\mathrm{con}%
}(k_{F}^{(g)})$ and the Fano factor by \cite{lafarge93} 
\begin{equation}
F_{\mathrm{dif}}(k_{F}^{(g)})=\frac{2R_{\mathrm{con}}^{2}F_{\mathrm{con}}+R_{%
\mathrm{sheet}}^{2}F_{\mathrm{sheet}}}{(2R_{\mathrm{con}}+R_{\mathrm{sheet}%
})^{2}},  \label{Fseries}
\end{equation}%
where $R_{\mathrm{sheet}}$ and $F_{\mathrm{sheet}}$ are respectively the
resistance and the Fano factor of the sheet, and $R_{\mathrm{con}}=G_{%
\mathrm{con}}^{-1}$ the single contact resistance. Theory of the anomalous
diffusion through the electron-hole puddles sea formed in graphene predicts
a universal scale-independent Fano factor $F_{\mathrm{sheet}}=1/3$ \cite%
{groth08} in agreement with a recent experiment \cite{dicarlo08}. In
addition numerical studies also indicate a nearly density independent Fano
factor (not universally equal to $1/3$) from moderate to strong disorder 
\cite{sanjose07,lewenkopf08}. According to Eq. (\ref{Fseries}), one expects $%
F_{\mathrm{dif}}(k_{F}^{(g)})=F_{\mathrm{sheet}}=1/3$ only when $R_{\mathrm{%
sheet}}\gg R_{\mathrm{con}}$ which might be the case in all devices measured
in Ref.~\cite{dicarlo08}. In contrast, when $R_{\mathrm{sheet}}\ll R_{%
\mathrm{con}}$, $F_{\mathrm{dif}}(k_{F}^{(g)})$ depends more on $F_{\mathrm{%
con}}$ which makes it decrease as a function of $|k_{F}^{(g)}|$.
Interestingly another experiment reports on such a decrease from $F_{\mathrm{%
dif}}(k_{F}^{(g)}=0)=1/3$ to a lower value at large densities \cite%
{danneau08}. Nevertheless it should be emphasized that the incoherent theory
underlying Eq.(\ref{Fseries}) is not valid for the very short graphene
devices investigated in Ref. \cite{danneau08}. It is thus necessary to
consider the nonlocal transport properties of the whole device.
Unfortunately the precise profile of the potential in this experiment is not
known and probably corresponds to $L\sim d$ making neither our study of the
ballistic Fabry-P\'{e}rot (Fig.~\ref{figureFbal}) nor the square well model
of Ref. \cite{tworzydlo06} quantitatively relevant. Nevertheless the
qualitative behavior of these experiments is captured by these models:
intrinsic noise close to the Dirac point and contact-dominated noise at
larger densities.

In conclusion, we have considered Dirac fermions scattering from one or two
potential steps having each a characteristic length $d$. Such steps
introduce additional dissipation localized at the source and drain of gFETs
and also modify drastically the noise properties of such devices. In the
ballistic regime, we predict that the presence of metallic contacts can
shift the conductance minimum to a non-zero charge density, which is
negative (resp. positive) for a metal which dopes the graphene with
electrons (resp. with holes). In addition we also suggest performing
conductance and noise measurements on a suspended bipolar graphene Fabry-P%
\'{e}rot structure \cite{bolotin08,andrei08,young08} with low doping at the
electrodes in order to observe enhanced Fabry-P\'{e}rot oscillations.
Finally, the asymmetry between electron and hole transport is enhanced when
these potential steps rise over longer distances.

\appendix

\section{Transmission across a potential step}

We consider the differential equation Eq.(\ref{f1}):%
\begin{equation}
f_{\alpha }^{\prime \prime }+(k_{F}^{2}(x)-k_{y}^{2}-\alpha ik_{F}^{^{\prime
}}(x))f_{\alpha }=0,
\end{equation}%
where 
\begin{equation}
k_{F}(x)=k_{F}^{(m)}+\frac{k_{F}^{(g)}-k_{F}^{(m)}}{e^{-x/d}+1}.
\end{equation}%
We introduce the new independent variable $\xi =-e^{-x/d}$ and seek for the
solutions in the general form $f_{\alpha }(x)=\xi ^{\mu }(1-\xi )^{-\lambda
}w_{\alpha }(\xi )$. The function $w_{\alpha }(\xi )$ satisfies the Gauss
hypergeometrical equation:%
\begin{equation}
\xi (1-\xi )w_{\alpha }^{\prime \prime }+\left[ c-(a+b+1)\xi \right]
w_{\alpha }^{\prime }-abw_{\alpha }=0,  \label{hyper}
\end{equation}%
if the exponents are chosen as 
\begin{equation}
\mu =-ik_{x}^{(g)}d\text{ \ \ \ and \ \ }\lambda
=-i(k_{F}^{(g)}-k_{F}^{(m)})d.
\end{equation}%
Introducing $\nu =ik_{x}^{(m)}d$, the parameters $a,b,c$ are given by 
\begin{eqnarray}
c &=&1+2\mu ,\text{ \ } \\
a+b+1 &=&1+2\mu -2\lambda , \\
a &=&(\mu -\lambda +\nu ), \\
b &=&(\mu -\lambda -\nu ).
\end{eqnarray}%
In the region $x\rightarrow +\infty $, namely $\xi \rightarrow 0$, the two
independent solutions of the hypergeometric equation are $w_{\alpha }(\xi
)=F(a,b,c;\xi )$ and $w_{\alpha }(\xi )=\xi ^{1-c}F(a-c+1,b-c+1,2-c;\xi )$ 
\cite{AbraStegun}. The corresponding functions $f_{\alpha }(x)=\xi ^{\mu
}w_{\alpha }(\xi )$ are the plane waves $\xi ^{\mu }=e^{ik_{x}^{(g)}d}$ and $%
\xi ^{-\mu }=e^{-ik_{x}^{(g)}d}$. We now construct a scattering state
containing a single oscillatory outgoing wave in the region $x\rightarrow
+\infty $, namely 
\begin{equation}
f_{\alpha }(x)\sim e^{ik_{x}^{(g)}x}\mbox{ at }x\rightarrow +\infty ,
\end{equation}%
where $k_{x}^{(g)}=s_{g}\sqrt{(k_{F}^{(g)})^{2}-k_{y}^{2}}$ is the
longitudinal momentum whose direction depends on the band index $s_{g}=%
\mathrm{sign}(k_{F}^{(g)})$ far on the right side. Note that although the
sign of $k_{x}^{(g)}$ can be either positive or negative depending on
doping, the group velocity always describes a right-moving charge.

From the general relation\cite{AbraStegun} between the hypergeometric
functions of respective arguments $\xi $ and $1/\xi $, one can extract the
asympotic behavior of $w(\xi )$ in the region for $\xi \rightarrow -\infty $ 
\begin{eqnarray}
w_{\alpha }(\xi ) &=&\frac{\Gamma (c)\Gamma (b-a)}{\Gamma (b)\Gamma (c-a)}%
(-\xi )^{-a} \\
&&+\frac{\Gamma (c)\Gamma (a-b)}{\Gamma (a)\Gamma (c-b)}(-\xi )^{-b}.
\end{eqnarray}%
Consequently the structure of the wave $f_{\alpha }(x)=\xi ^{\mu }(1-\xi
)^{-\lambda }w(\xi )\sim \xi ^{\mu }(-\xi )^{-\lambda }w(\xi )$ in the
region $x\rightarrow -\infty $ (namely $\xi \rightarrow -\infty $) consists
in two parts 
\begin{equation}
f_{\alpha }(x)\sim f_{\alpha }^{(inc)}(x)+f_{\alpha }^{(ref)}(x)\mbox{ at }%
x\rightarrow -\infty ,
\end{equation}%
which are respectively the incident wave%
\begin{equation}
f_{\alpha }^{(inc)}(x)=(-1)^{\mu }\frac{\Gamma (c)\Gamma (b-a)}{\Gamma
(b)\Gamma (c-a)}e^{ik_{x}^{(m)}x},
\end{equation}%
and the reflected wave 
\begin{equation}
f_{\alpha }^{(ref)}=(-1)^{\mu }\frac{\Gamma (c)\Gamma (a-b)}{\Gamma
(a)\Gamma (c-b)}e^{-ik_{x}^{(m)}x}.
\end{equation}%
We have used 
\begin{eqnarray}
(-\xi )^{-a+\mu -\lambda } &=&(e^{-x/d})^{-\nu }=e^{ik_{x}^{(m)}x} \\
(-\xi )^{-b+\mu -\lambda } &=&(e^{-x/d})^{\nu }=e^{-ik_{x}^{(m)}x}
\end{eqnarray}%
We emphasize again that $e^{ik_{x}^{(m)}x}$ is always the right-moving
incident wave, although $k_{x}^{(m)}$ can be either positive or negative
(because the projection of the group velocity is positive in both n-type and
p-type doped graphene). Therefore the reflection probability is:%
\begin{eqnarray}
\mathfrak{R}_{\mathrm{step}} &=&\left\vert \frac{\Gamma (b)\Gamma (c-a)}{%
\Gamma (a)\Gamma (c-b)}\right\vert ^{2} \\
&=&\left\vert \frac{\Gamma (\mu -\lambda -\nu )\Gamma (1+\lambda +\mu -\nu )%
}{\Gamma (\mu -\lambda +\nu )\Gamma (1+\lambda +\mu +\nu )}\right\vert ^{2}
\end{eqnarray}%
which leads to Eqs.(\ref{RGamma},\ref{R1}) in the text.

\section{Wood-Saxon potential}

The Wood-Saxon potential corresponds to two symmetric steps 
\begin{eqnarray}
k_{F}(x) &=&k_{F}^{(m)}+\left( k_{F}^{(g)}-k_{F}^{(m)}\right) \times  \notag
\\
&&\left( \frac{\theta (-x)}{e^{-(x+L/2)/d}+1}+\frac{\theta (x)}{%
e^{(x-L/2)/d}+1}\right)  \label{wood-saxon}
\end{eqnarray}%
We have checked that the transmission probability is given by the formula of
Ref. \cite{kennedy02}, namely $\mathcal{T}_{\mathrm{ball}}(k_{y})=\left\vert
Ae^{\mu L/d}/(1-Ce^{2\mu L/d})\right\vert ^{2}$ where 
\begin{eqnarray}
A &=&\left( \frac{(\mu +\nu )^{2}-\lambda ^{2}}{4\mu \nu }\right) \\
&&\frac{\Gamma ^{2}(-\mu -\nu -\lambda )\Gamma ^{2}(-\mu -\nu +\lambda )}{%
\Gamma ^{2}(-2\mu )\Gamma ^{2}(-2\nu )}, \\
C &=&\left[ \frac{(\mu +\nu )^{2}-\lambda ^{2}}{(\mu -\nu )^{2}-\lambda ^{2}}%
\right] \frac{B^{2}(2\mu ,-\mu -\nu +\lambda )}{B^{2}(-2\mu ,\mu -\nu
-\lambda )}.
\end{eqnarray}%
We denote $B(a,b)=\Gamma (a)\Gamma (b)/\Gamma (a+b)$ the beta function, and $%
\mu =-ik_{x}^{(g)}d,$ \ $\nu =ik_{x}^{(m)}d$, $\lambda
=-i(k_{F}^{(g)}-k_{F}^{(m)})d$ like in appendix A.

\begin{acknowledgments}
We are very grateful to Bjoern Trauzettel, Alexander Buzdin and Nimrod
Stander for carefully reading of the manuscript. J.C. acknowledges the
Geballe Laboratory for Advanced Materials at Stanford for hospitality during
the completion of this work. This work was supported by the Institut
Universitaire de France (Chair of A. Buzdin), the Agence Nationale de la
Recherche under grant ANR-07-NANO-011-05 (ELEC-EPR), the MARCO/FENA program
and the Air Force Office of Scientific Research.
\end{acknowledgments}


\begin{thebibliography}{99}
\bibitem{footnote} In the whole paper, we take the formal limit $%
W/L\rightarrow \infty $ and express the results in terms of conductance per
channel, $W$ and $L$ being respectively the width (along $y$) and the length
(along $x$) of the graphene strip. In practice, this corresponds to samples
with aspect ratio $W/L\gtrsim 4$ according to Ref.\cite{tworzydlo06}.
\end{thebibliography}
\end{document}